\documentclass[aps,twocolumn,showpacs,floatfix,amssymb]{revtex4}
\usepackage{graphicx}
\begin{document}

\title{Spin excitations and mechanisms of superconductivity
in cuprates}

\author{  N. M. Plakida\footnote{E-mail: plakida@theor.jinr.ru} }
\affiliation{Joint Institute for Nuclear Research, 141980 Dubna, Russia}

\date{\today}

\begin{abstract}
A microscopic theory of spin excitations in strongly-correlated
electronic systems within the  $t$-$J$ model is discussed. An
exact representation for the  dynamic spin susceptibility  is
derived. In the normal state, the excitation spectrum  reveals a
crossover from spin-wave-like excitations at low doping to
overdamped paramagnons above the optimal doping. At low
temperatures, the resonance mode  at the antiferromagnetic wave
vector ${\bf Q} = \pi(1,1)$ emerges  which is explained by a
strong suppression of the spin excitation damping  caused by a
spin gap at ${\bf Q}$ rather than by opening of a superconducting
gap. A major role of spin excitations in the $d$-wave
superconducting pairing in cuprates is stressed in discussing
mechanisms of high-$T_c$ superconductivity within the Hubbard
model in the limit of strong correlations, while electron-phonon
interaction and a well-screened weak Coulomb interaction are not
essential.
\end{abstract}


\maketitle




\section{Introduction}
\label{intro}

Recent studies of  electron  and spin-excitation spectra using
angle-resolved photoemission (ARPES) and inelastic neutron
scattering (INS) have revealed an important role of
antiferromagnetic (AF) spin excitations in the ``kink''
phenomenon and the $d$-wave pairing in cuprates. In particular,
in Ref.~\cite{Dahm09} quantitative analysis of the AF
spin-excitation spectrum  measured by INS and of ARPES data for
the spin-fermion coupling of the same YBa$_2$Cu$_3$O$_{6.6}$
(YBCO$_{6.6}$) crystal were used for numerical solution of the
Eliashberg-type equations. The superconducting transition
temperature  was found to exceed  $T_c = 150$~K.

The main argument against the spin-fluctuation pairing mechanism,
a weak intensity of spin fluctuations at the optimal doping seen
in  INS experiments~\cite{Bourges98}, was dismissed in recent
resonant inelastic x-ray scattering~\cite{LeTacon11}. In a large
family of cuprates  paramagnon AF excitations with dispersions
and spectral weights similar to those of magnons in undoped
cuprates were found. A numerical solution of the Eliashberg
equations for the magnetic spectrum found in YBCO$_7$ and for the
electron-spin interaction described by the $t$--$J$ model results
in $T_c = 100-200$~K. These calculations based on experimental
data demonstrate that spin fluctuations have sufficient strength
to mediate high-temperature superconductivity in cuprates and,
therefore, alternative mechanism based on electron-phonon
interaction (EPI) (see, e.g.,~\cite{Maksimov10}) seems to play a
secondary role in cuprate superconductivity. Strong EPI observed
in polaronic effects in cuprates may be irrelevant for the
$d$-wave pairing mediated  by $l=2$ component of EPI as pointed
out in Ref.~\cite{Plakida11}.

In this report we briefly consider a microscopic theory of
spin-excitation spectrum in strongly correlated electronic
systems (SCES)~\cite{Vladimirov09,Vladimirov11}. Using a model
for the spin-excitation spectrum, we consider spin-fluctuation
pairing within the Hubbard model in the limit of strong
correlations, $U \gg t$~\cite{Plakida03,Plakida07}. To compare
various mechanisms of superconducting $d$-wave pairing, we take
into account also EPI and a well-screened weak Coulomb
interaction considered in Ref.~\cite{Alexandrov11}. We show that
the latter gives a small contribution for the $d$-wave pairing
and cannot suppress the superconductivity.

\section{Spin-excitation spectrum}
\label{sec:1}

To describe the low-energy spin excitations  in SCES  the
one-subband $t$--$J$ model can be used:
\begin{equation}
H= \sum_{i\neq j,\sigma} \, t_{ij}\, \hat{c}_{i\sigma}^{\dag}
\hat{c}_{j\sigma}
 + \frac{1}{2} \sum_{i\neq j} \, J_{ij}\;
 ( {\bf S}_{i}{\bf S}_{j} - {\frac {n_{i} n_{j}}{4}} ) ,
\label{b1}
\end{equation}
where $t_{ij}$ is the hopping integral and $J_{ij}$ is the
exchange interaction. Here $ \hat{c}_{i\sigma}^{\dag} =
c_{i\sigma}^{\dag} \, (1 - n_{i, -\sigma})\,$ are the projected
Fermi operators acting in the the singly occupied subband and
$n_{i} = \sum_{\sigma} n_{i,\sigma}, \, n_{i,\sigma} =
\hat{c}_{i\sigma}^{\dag} \hat{c}_{i\sigma}$.  $\,
 S^{\alpha}_{i} = (1/2)\sum_{\sigma \sigma'} \hat{c}_{i\sigma}^{\dag}
\tau^\alpha_{\sigma \sigma'}\hat{c}_{i\sigma'}$ are the
spin-$1/2$ operators where $\tau^\alpha_{\sigma \sigma'}$ are the
Pauli matrices, $\sigma = \pm 1$.
\par
Using the  projection technique for the Kubo-Mori relaxation
functions,  an exact representation for the dynamical spin
susceptibility (DSS),  the retarded Green function (GF) of  the
transverse spin-density operators $\,S_{\bf q}^{\pm} = S^{x}_{\bf
q} \pm i S^{y}_{\bf q}\,$, can be derived~\cite{Vladimirov09}
(see also~\cite{Sega03}):
\begin{equation}
\chi({\bf q}, \omega)=  -\langle \!\langle {S}_{\bf q}^{+}|
{S}_{-\bf q}^{-} \rangle \!\rangle_{\omega} =
  \frac{m({\bf q})} { \omega_{\bf q}^2 +\omega \, \Sigma({\bf
q},\omega) - \omega^2 },
 \label{b2}
\end{equation}
where $m({\bf q})=\langle [i\dot{S}^{+}_{\bf q}, S_{-\bf
q}^{-}]\rangle  = \langle [\, [{S}^{+}_{\bf q}, H], \, S_{-\bf
q}^{-}]\rangle$. The static spin-excitation spectrum $\omega_{\bf
q}$ is calculated from the equality for Kubo-Mori correlation
function $\, m({\bf q}) = (-\ddot{S}_{\bf q}^{+},S_{-{\bf
q}}^{-}) =  \omega_{\bf q}^2 \, ({S}_{\bf q}^{+},S_{-{\bf
q}}^{-}), \,$ where  $(-\ddot{S}_{\bf q}^{+},S_{-{\bf q}}^{-})$ is
evaluated in a generalized mean-field
approximation~\cite{Vladimirov09}. The self-energy is given by
the retarded GF,
\begin{equation}
\Sigma({\bf q},\omega) = [1/ m({\bf q})\,\omega]\, \langle
\!\langle - \ddot{S}_{\bf q}^{+}\,| - \ddot{S}_{-\bf
q}^{-}\rangle \!\rangle_{\omega}^{(\rm pp)} .
 \label{b3}
\end{equation}
The ``proper part'' (\rm pp) of the GF(\ref{b3})  describes  the
projected time evolution as in the original Mori projection
technique. The self-energy (\ref{b3}) is defined in terms of  the
force operators $\,  -\ddot{S}_{i }^{\pm} =[[{S}_{i}^{\pm},
(H_{t} + H_{J})] , (H_{t} +H_{J} )] \equiv \sum_{\alpha}
F_{i}^{\alpha}$ ($\alpha = tt,\, tJ,\, Jt, \, JJ$), where $H_{t}$
and  $H_{J}$ are the hopping  and the exchange parts of the
Hamiltonian~(\ref {b1}).
\par
In the Heisenberg limit at zero doping, $\delta = 0$, the
self-energy is determined by the force $ F_{i}^{JJ}$. At a finite
hole doping, $\delta > 0.05$, the largest contribution to the
self-energy (\ref {b3}) is given by the hopping term $ F_{i}^{tt}
= \sum_{j,n}t_{ij}\Bigl\{
t_{jn}\left[H^{-}_{ijn}+H^{+}_{nji}\right]
  -  (i \Longleftrightarrow j )\Bigr\}, \,$
where  $\,H^{-}_{ijn}  = \hat{c}_{i\sigma}^{\dag}
S_{j}^{-}\hat{c}_{n\sigma} + \hat{c}_{i\downarrow}^{\dag} (1-
n_{j, - \sigma} )\hat{c}_{n \uparrow}\,$.  We calculate the
self-energy in the mode-coupling approximation (MCA), $\,
\langle \hat{c}_{i\sigma}^{\dag} S_{j}^{-}\hat{c}_{n\sigma}|
\hat{c}_{n'\sigma}^{\dag}(t)
S_{j'}^{+}(t)\hat{c}_{i'\sigma}(t)\rangle =   \langle
\hat{c}_{i\sigma}^{\dag}\hat{c}_{i'\sigma}(t) \rangle $\, $
\langle S_{j}^{-}S_{j'}^{+}(t)\rangle $\, $ \langle
\hat{c}_{n\sigma} \hat{c}_{n'\sigma}^{\dag}(t)\rangle \,$.  In
the superconducting state,   the anomalous
correlation functions  $ \langle \hat{c}_{i, -
\sigma}^{\dag}\hat{c}_{n'\sigma}^{\dag}(t)\rangle\;$ $\langle
S_{j}^{-}S_{j'}^{+}(t)\rangle $ $ \langle \hat{c}_{n\sigma}
\hat{c}_{i', -\sigma}(t) \rangle \,$ are also taken into account.
Using the spectral representation for these two-time correlation
functions  both the real, $\Sigma'({\bf q},\omega)$, and the
imaginary, $\Sigma''({\bf q},\omega)$, parts of the self-energy
(\ref{b3}) are calculated~\cite{Vladimirov11}.
\par
The spectrum of spin excitations $\omega_{\bf q}$ and the damping
$\, \Gamma_{\bf q}= - (1/2)\Sigma''({\bf q},\omega_{\bf q})\,$
are calculated in a broad region of temperature and doping. In
the Heisenberg limit at $\delta = 0$ the spectrum of spin
excitations reveals well-defined quasiparticles  with
$\Gamma_{\bf q} \ll \omega_{\bf q}$ characteristic to the
Heisenberg model. However, for non-zero doping the spin-electron
scattering contribution $\,\Sigma''_{t}({\bf q},\omega)\,$
increases rapidly with doping and temperature and already at
moderate hole concentration far exceeds the spin-spin scattering
contribution $\,\Sigma''_{J}({\bf q},\omega)\,$. We conclude,
that at low enough doping and low temperatures well-defined
spin-wave-like excitations propagating on the AF short-range
order background are observed, while  for higher doping and
temperatures a crossover to AF paramagnon-like spin  excitations
occurs as found in INS experiments.
\begin{figure}\centering
\includegraphics[width=0.35\textwidth]{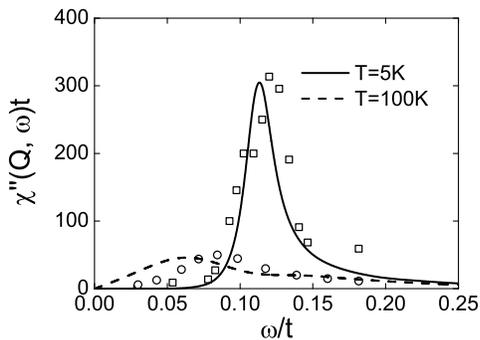}
\caption{Spectral function  at $\delta = 0.2$ compared with
experimental data~\cite{Bourges98} at $T=5K$ (squares) and
$T=100K$ (circles).}
 \label{fig1}
\end{figure}
\begin{figure}\centering
\includegraphics[width=0.35\textwidth]{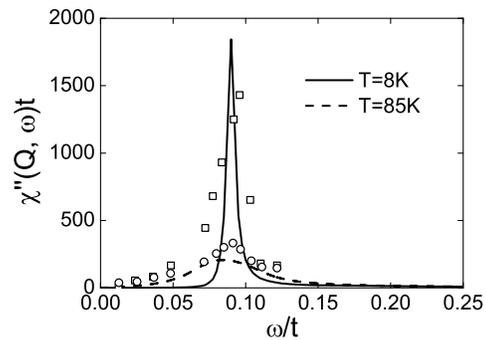}
\caption{Spectral function at $\delta = 0.09$ compared with
experimental data~\cite{Stock04}  at $T=8K$ (squares) and $T=85K$
(circles).}
 \label{fig2}
\end{figure}
In the superconducting state the spectral function $\chi''({\bf
Q}, \omega) = {\rm Im}\,\chi({\bf Q}, \omega)$ were calculated
assuming the $d$-wave gap function $\Delta_{\bf q} =
(\Delta/2)(\cos q_x - \cos q_y)$~\cite{Vladimirov11}. The DSS
(\ref{b2}) reveals a pronounced resonance mode (RM) at low
temperatures  due to a strong suppression of the damping of spin
excitations. This is explained by an involvement of a spin
excitation  in the decay process described by creation of three
excitations: particle-hole pair with energies $\,\omega_1 +
\omega_2\,$ and a spin excitation with energy $\omega_3$  which is
controlled by the energy and momentum conservation laws, $\omega
= \omega_1 + \omega_2 + \omega_3\,$ and $\, {\bf q} = {\bf q}_1 +
{\bf q}_2 + {\bf q}_3 $.  Due to the spin gap in the
spin-excitation spectrum at ${\bf Q}$ the spin excitation with
the energy $\omega_3 \simeq \omega_{\bf Q}$ in this process plays
a dominant role in limiting the decay of the RM in comparison
with the superconducting gap in the particle-hole excitation.
Since $\omega_{\bf Q}$ shows a weak temperature dependence at $ T
\lesssim T_c $ the RM does not reveal an appreciable temperature
dependence and can be observed even above $T_c$ in the underdoped
region (see, e.g.,~\cite{Stock04,Hinkov07}).
\par
Figure~\ref{fig1} shows  the temperature dependence of the
spectral functions  in the overdoped case at $\delta = 0.2$ and
experimental data (symbols) for YBCO$_{6.92}$~\cite{Bourges98}.
The RM having a high intensity at low temperatures strongly
decreases with temperature and becomes very broad at $T \sim
T_{\rm c}$. The spectral function  for the underdoped case
$\delta = 0.09$ is plotted Fig.~\ref{fig2}. The RM shows a weak
temperature  dependence and is still visible even at $T = 85$~K$=
1.4 \,T_{\rm c}$ as found in YBCO$_{6.5}$ crystal~\cite{Stock04}.

Thus, as compared with the spin-exciton scenario for the RM based
on the random-phase approximation where only electron-hole bubble
diagrams are taken into  account (see, e.g.,~\cite{Sega03}), we
propose an alternative explanation of the RM which is driven by
the spin gap at ${\bf Q}$ rather than by  opening of the
superconducting gap.

\section{Spin-fluctuation $d$-wave pairing}
\label{sec:2}

Despite of intensive search for the mechanism of high-temperature
superconductivity in cuprates, there is still no commonly
accepted theory (for a review see~\cite{Plakida10}). A
microscopic theory of superconducting $d$-wave  pairing mediated
by  AF exchange interaction and spin-fluctuations induced by
kinematic interaction has been developed within the  $t$--$J$
model in Ref.~\cite{Plakida99} and the Hubbard model in
Ref.~\cite{Plakida03}.

Recently, the problem of superconductivity in the repulsive
Hubbard model in the weak correlation limit was discussed. In
Ref.~\cite{Raghu10} an asymptotically exact  solution for the
$d$-wave pairing was found, while consideration of a
well-screened weak Coulomb interaction (CI) has not shown a
possibility for superconducting pairing~\cite{Alexandrov11}. To
resolve this controversy, we have considered superconductivity in
the Hubbard model in the limit of strong correlations, $U \gg t$,
taking into account also a well-screened weak CI and EPI:
\begin{eqnarray}
&&H= \varepsilon_1\sum_{i,\sigma}X_{i}^{\sigma \sigma} +
\varepsilon_2\sum_{i}X_{i}^{22} +  \sum_{i\neq
j,\sigma}\,t_{ij}\, \bigl\{ X_{i}^{\sigma 0}X_{j}^{0\sigma}
 \nonumber \\
&& + X_{i}^{2 \sigma}X_{j}^{\sigma 2} + 2 \sigma
(X_{i}^{2\bar\sigma}X_{j}^{0 \sigma} + {\rm H.c.})\bigr\} + H_{c,
ep}.
 \label{2}
\end{eqnarray}
We introduced here the Hubbard operators (HOs)
$X_{i}^{\alpha\beta} = |i\alpha\rangle\langle i\beta|$  for the
four states on the lattice site $i$: an empty state $\,\alpha
=|0\rangle$, a one-hole state $\,\alpha =|\sigma\rangle$ with the
spin $\sigma = \pm 1/2 = (\uparrow,\downarrow),\, \bar{\sigma} =
-\sigma$,  and  a two-hole state $\,|2\rangle =|\uparrow
\downarrow \rangle $. To apply the model for cuprate
superconductors, we introduce the single-particle energy
$\varepsilon_1=\varepsilon_d-\mu$ as an energy of the one-hole
$d$-state.  The two-hole energy $\varepsilon_2 = 2\varepsilon_1+ U
$ is an energy of the $p$-$d$ singlet state  where   $ U =
 \epsilon_p-\epsilon_d$ is the charge-transfer
energy between the oxygen $p$ and copper $d$ states.
\par
The last term in (\ref{2}) denotes a weak screened CI $V(ij)$
between charge carriers in the plane and  EPI $g(ij)$ for charge
carriers
\begin{eqnarray}
H_{c, ep} &= & \frac{1}{2}  \sum_{i\neq j}\,V(ij) N_i N_j +
\sum_{i, j}\,g(i,j) N_i\, u_j ,
 \label{3}
\end{eqnarray}
where $u_j$ is a displacement for a particular phonon mode. $N_i =
\sum_{\sigma} X_{i}^{\sigma \sigma} + 2 X_{i}^{22}\,$  is the
number operator. The chemical potential $\mu$ depends on the
average  hole occupation number $\, n =  1 + \delta = \langle\,
N_i \rangle $.
\par
Using the projection technique in the equation of motion method
for the GF in terms of the HOs as described in
\cite{Plakida03,Plakida07} we can derive an exact Dyson equations
for the two-subband  matrix GFs. The normal GF can be written as,
\begin{eqnarray}
&&  {\hat G}({\bf k},\omega) = \langle\! \langle \left(
\begin{array}{c}
     X_{{\bf k}}^{\sigma2}  \\
     X_{{\bf k}}^{0 \bar\sigma } \\
        \end{array}\right)  \mid
 ( X_{{\bf k}}^{2\sigma}\,  X_{{\bf k}}^{\bar\sigma 0}\,
 \rangle \! \rangle_{\omega}
\nonumber  \\
&&   =  \Bigl(
  \hat {G}^{-1}_{N}({\bf k},\omega)
  +   \hat{\varphi}_\sigma({\bf k},\omega)\,
  \hat{G}_{N}({\bf k},- \omega)\,\hat{\varphi}^{*}_\sigma({\bf
k},\omega)  \Bigr)^{-1}  \hat{Q},
 \nonumber  \\
 && \quad
{\hat G}_{N}({\bf k},\omega)= \Bigl( \omega \hat\tau_0 -
\hat{\varepsilon}({\bf k}) -
  \hat{\Sigma}({\bf k},\omega) \Bigr)^{-1},
\label{4}
 \end{eqnarray}
where $\hat{\varepsilon}({\bf k})$ is the hole energy in the
mean-field approximation (MFA) and $\hat{\Sigma}({\bf k},\omega)$
is the normal self-energy. The anomalous (pair) GF reads,
\begin{eqnarray}
&&   {\hat F}_\sigma({\bf k},\omega)= \langle\! \langle \left(
\begin{array}{c}
     X_{\bf k}^{\sigma2}  \\
     X_{\bf k}^{0 \bar\sigma } \\
        \end{array}\right)  \mid
 ( X_{-\bf k}^{\bar\sigma 2}\, \, X_{-\bf k}^{0\sigma})
 \rangle \! \rangle_{\omega}
\nonumber  \\
 &&  =  -\hat{G}_{N}({\bf k},-
\omega)\,\hat{\varphi}_\sigma({\bf k},\omega) \,
 \hat{G}_\sigma({\bf k},\omega).
 \label{5}
\end{eqnarray}
The superconducting gap function  $\, {\hat \varphi}_\sigma({\bf
k},\omega) = \hat{\Delta}_{\sigma}({\bf k}) +
 \hat\Phi_{\sigma}({\bf k},\omega)
$ has a nonretarded contribution $\hat{\Delta}_{\sigma}({\bf k})$
determined by the AF exchange interaction and CI in MFA and the
anomalous self-energy $\hat\Phi_{\sigma}({\bf k},\omega)$.
\par
The self-energies $\hat{\Sigma}({\bf k},\omega),\,
\hat\Phi_{\sigma}({\bf k},\omega)$ are calculated in the MCA by
assuming an independent propagation of Fermi-like and Bose-like
excitations in multiparticle GFs. Below we consider the
hole-doped case, $n = 1 + \delta
> 1$. The diagonal components of the self-energies for the
two-hole subband can be written as
\begin{eqnarray}
  \Sigma^{22}({\bf k},\omega) & = &
   \frac{1}{N} \sum\sb{\bf q}
   \int\limits\sb{-\infty}\sp{+\infty} {\rm d}z \,
   K^{(+)}(\omega,z|{\bf q },{\bf k-q})
   \nonumber \\
& \times & [- ({1}/{\pi Q_2})\, \mbox{Im}\, G\sp{22}({\bf q},z)] ,
 \label{6}
\end{eqnarray}
\begin{eqnarray}
 \Phi^{22}_{\sigma}({\bf k},\omega) & = &
 \frac{1}{N} \sum_{\bf q} \int\limits\sb{-\infty}\sp{+\infty} {\rm d}z \,
   K^{(-)}(\omega,z|{\bf q },{\bf k-q})
  \nonumber \\
&& \times  [- ({1}/{\pi Q_2}) \,\mbox{Im}\,
F^{22}\sb{\sigma}({\bf q},z)].
 \label{7}
\end{eqnarray}
where  $Q_2 = n/2$ is the weight of the second subband. The kernel
of these integral equations has a form, similar to the
strong-coupling Eliashberg theory~\cite{Eliashberg60}:
\begin{eqnarray}
 &&   K^{(\pm)}(\omega,z |{\bf q },{\bf k-q })  =
\int\limits\sb{-\infty}\sp{+\infty}\frac{d \omega'}{2\pi}
 \;  \frac{\tanh \frac{z}{2T} +
 \coth \frac{\omega'}{2T}}{\omega - z - \omega'}\,
   \nonumber \\
&&\Big\{ |t({\bf q})|^{2} {\rm Im}\, \chi\sb{sf}({\bf k-
q},\omega') \pm |g_{{\bf k -q}}|^2 {\rm Im}\, \chi_{ph}({\bf k-q},
\omega')
 \nonumber \\
&&  \pm \left[ |V_{{\bf k -q}}|^2 + | t({\bf q})|^{2}/4 \right]
 {\rm Im}\, \chi_{cf}({\bf k-q}, \omega')\Big\},
  \label{8}
\end{eqnarray}
where the  spectral density of bosonic excitations are determined
by the dynamic susceptibility for spin fluctuations, $\,
\chi\sb{sf}({\bf q},\omega)=
  -  \langle\!\langle {\bf S\sb{q} | S\sb{-q}}
\rangle\!\rangle\sb{\omega}$, charge fluctuations $
\chi\sb{cf}({\bf q},\omega) =
 - \langle\!\langle  N\sb{\bf q} |  N\sb{-\bf q}
   \rangle\!\rangle\sb{\omega}$, and phonon GF $\chi\sb{ph}({\bf q},\omega) =
 - \langle\!\langle  u\sb{\bf q} | u\sb{-\bf q}
   \rangle\!\rangle\sb{\omega}$.
The gap equation takes the form:
\begin{eqnarray}
&& \varphi_{2,\sigma}({\bf k},\omega)=
 \frac{1}{N} \sum_{\bf q}
  \int\limits\sb{-\infty}\sp{+\infty} dz
  \Big\{[ J_{\bf k-q}- V_{\bf k-q}]\,
  \frac{1}{2}\tanh\frac{z}{2T}
\nonumber \\
&&  +  \; K^{(-)}(\omega,z|{\bf q },{\bf k-q})
  \Big\}
 [- ({1}/{\pi}Q_2) \,\mbox{Im}\,  F^{22}\sb{\sigma}({\bf q},z)].
\label{9}
\end{eqnarray}
Here the exchange interaction  $J_{\bf q} = 2 J\,(\cos q_x +\cos
q_y)$ induces  pairing in MFA, while the Coulomb repulsion
$V_{\bf k-q}$ suppresses  the pairing. The  pairing induced by
retarded interactions is described by the kernel  (\ref{8}).
\par
To estimate contributions from various interactions in the gap
equation  (\ref{9}) we consider a weak coupling approximation for
the kernel  (\ref{8}), $ K^{(-)}(\omega, z|{\bf q },{\bf q'})
\simeq K^{(-)}(\omega = 0, z =0 |{\bf q },{\bf q'}) $. In this
approximation the gap equation reduces to the BCS-type form where
the interactions are determined by the  static susceptibility,
$\chi_{\bf q } = (1/\pi) \int_{-\infty}^{+\infty}
(d\omega/\omega) {\rm Im}\, \chi({\bf q}, \omega)$:
\begin{eqnarray}
&& \varphi({\bf k})=
 \frac{1}{N} \sum_{\bf q}\Big\{
     J_{\bf k-q}- V_{\bf k-q}
-  |t({\bf q})|\sp{2}\; \chi_{sf}({\bf k -q})
 \nonumber \\
&& + |g_{{\bf k -q}}|^2 \chi_{ph}({\bf k-q})
\Big\}\,\frac{\varphi({\bf q})}{2 E_{\bf q}} \tanh \frac{E_{\bf
q}}{2T},
     \label{10}
\end{eqnarray}
where $ E_{\bf q} = [\varepsilon^2_{\bf q} + |\varphi({\bf
q})|^2]^{1/2}$ and $\varphi({\bf k}) =\varphi_{2,\sigma}({\bf k},
0)$.  The unimportant  contribution from charge fluctuations
$\chi\sb{cf}({\bf k -q}) $ is omitted here (see later). To obtain
an equation for superconducting $T_c$ it is sufficient to
consider a linearized gap equation (\ref{10}). Using a model
$d$-wave gap function, $\varphi({\bf k}) = \Delta \,  \eta_{\bf k},\,
\eta_{\bf k} = (\cos k_x - \cos k_y)$, a linearized  gap equation
(\ref{10}) for $ T_c $ can be written as:
\begin{equation}
 1 = \frac{1}{N}\sum_{\bf q}  [ J - \widehat{V_c}  - |t({\bf q})|^{2}\,\widehat{\chi}\sb{sf}
 +   \widehat{V_{ep}}]\frac{\eta^2_{\bf q}}{2
\varepsilon_{\bf q}}\tanh\frac{\varepsilon_{\bf q}}{2T_c}\, .
 \label{11}
\end{equation}
The coupling constants  are given by the expressions:
\begin{eqnarray}
&&  \widehat{V_c} =\frac{1}{N}\sum_{\bf k}  V({\bf k})\, \cos k_x, \;
 \widehat{\chi\sb{sf}} =\frac{1}{N}\sum_{\bf k}
\chi\sb{sf}({\bf k})\, \cos k_x ,
 \nonumber \\
&&  \widehat{V\sb{ep}} =  \frac{1}{N}\sum_{\bf k}
   |g({\bf k })|^2\chi\sb{ph}({\bf k})\,\cos k_x .
    \label{12}
\end{eqnarray}
To estimate the contribution $\widehat{V_c}$ from the CI   we
consider a model for the 2D screened CI  suggested in
Ref.~\cite{Alexandrov11}:
\begin{equation}
  V({\bf k}) =   u_c \,\frac{1}{|{\bf k}| + \kappa },
 \;   u_c  =\frac{2\pi e^2}{a \, \varepsilon_{0}},
   \label{13}
\end{equation}
where $\kappa$  is the inverse  screening length ($|{\bf k}|$ and
$\kappa$ are measured in units of $1/a$),  $a$ is the lattice
constant, and $\varepsilon_{0}$ is the static dielectric constant
of the lattice (in cuprates $\varepsilon_{0} \sim 30$). For the
static spin-fluctuation susceptibility we adopt the model as
in~\cite{Plakida03,Plakida07}:
\begin{equation}
  \chi_{sf}(\xi, {\bf k}) =
 \frac {\chi_0}{1+ \xi^2 [1+ (1/2)(\cos k_x + \cos k_y)]}.
 \label{14}
\end{equation}
Here  $\chi_{0}= ({3}/{4 \omega_{s}})(1- \delta) [ ({1}/{N})
\sum_{\bf q} \chi_{sf}({\bf q})/\chi_{0} ]^{-1}$  is fixed by the
condition: $ \langle {\bf S}_{i}^2\rangle =(3/4)(1- \delta)$
where $\omega_{s} \sim J$ is a characteristic spin-excitation
energy. The EPI coupling
constant $\widehat{V\sb{ep}}$ strongly depends on the ${\bf
k}$-variation  of the EPI matrix element $|g({\bf k })|^2$ and a
phonon dispersion in $ \chi\sb{ph}({\bf k}) = 1/M\omega^2_{\bf k
}$. In particular, for a local interaction $g({\bf k }) = g$ and
a dispersionless optic phonon, $\omega_{\bf k } = \omega_0$ the
coupling constant for the $d$-wave pairing vanishes, $
\widehat{V\sb{ep}} = 0$. A large electron-phonon coupling for
the $d$-wave pairing can occur for a strong forward scattering,
$k \rightarrow 0$ in EPI (see, e.g.,~\cite{Maksimov10,Lichtenstein95}).
\par
Numerical integration in  (\ref{12})  for the  model  (\ref{13})
gives  for the CI coupling constant:
\begin{equation}
\widehat{ V_c} = u_c \, 0.05\;  (0.11),\quad \widehat{ V_{c}}/
\widehat{ V_{c0}} = 0.26\; (0.38),
 \label{15}
\end{equation}
for $\kappa =  1 \; (0.2) \,$, respectively. A small ratio
$\widehat{ V_{c}}/ \widehat{ V_{c0}}$, where $\widehat{V_{c0}} =
(1/N)\sum_{\bf k}  V({\bf k})\,$ shows that for the $d$-wave
pairing the repulsion induced by CI is remarkably  suppressed. In
particular, for $u_c \simeq 1~eV$ we have still a positive,
though a small contribution from the AF exchange interaction, $J -
\widehat{ V_c} = 0.08\, (0.02)$ for $J = 0.13$~eV. Therefore, in
MFA we obtain only a weak coupling and a low $T_c$ (cf. with
~\cite{Plakida03,Plakida99}).
\par
The spin-fluctuation coupling constant in  (\ref{12})  for the
model susceptibility (\ref{14}) is given by,
\begin{equation}
\widehat{\chi\sb{sf}}(\xi)
  = - 0.66,\;(-  0.26), \quad  \chi_0(\xi) = 14.8 \; (3.4) \, ,
 \label{16}
\end{equation}
in units of $1/t = 0.4 /\omega_s\,$ for $ \xi =3.4 \;(\xi = 1.4) $
at hole doping $\delta = 0.05 \, (0.30)$,
respectively~\cite{Vladimirov09}. While the spin susceptibility
$\chi_0 = \chi_{sf}({\bf Q}) $ at the AF wave vector ${\bf Q}$
is positive and quite large, the contribution of the static
susceptibility to the coupling constant
$\widehat{\chi\sb{sf}}(\xi)$  (\ref{16}) is negative that results
in attraction mediated by spin-fluctuations in the equation
(\ref{11}) for $T_c$. In the underdoped region with large AF
correlation length $\xi$ the spin-fluctuation  coupling constant
is quite large, while for the overdoped region with small $\xi$
the coupling reduces resulting in lowering  of $T_c$. Using a
conventional dispersion for electrons: $ t({\bf q}) = 2 t\,(\cos
q_x +\cos q_y) + 4 t' \,\cos q_x \cos q_y $ with $t = 0.4$~eV and
$ |t'/t| \sim 0.2 $, we can estimate the spin-fluctuation
coupling constant averaged over the Fermi surface, $\langle ...
\rangle_{\rm F}$ as: $ (1/t)\,\langle |t({\bf q})|^2\rangle_{\rm
F} \simeq 4\, t \simeq 1.6 $~eV.  Numerical estimation for the
charge fluctuation susceptibility appears negligibly small, $
\widehat{ \chi\sb{cf}} \sim (1/t) \times 10^{-3}$ which results
in a small contribution from the CI in the kernel (\ref{8}).
\begin{figure}\centering
\includegraphics[width=0.3\textwidth]{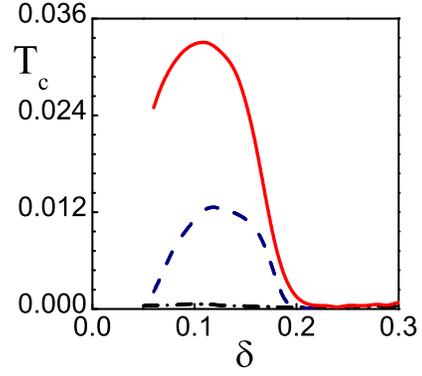}
\caption{(Color online) $T_c(\delta)$ (red solid line) compared
with pairing induced  by  spin fluctuations (blue dashed line)
and AF interaction  (dotted black line) (in units of $t$).}
 \label{fig3}
\end{figure}
\par
\begin{figure}\centering
\includegraphics[width=0.3\textwidth]{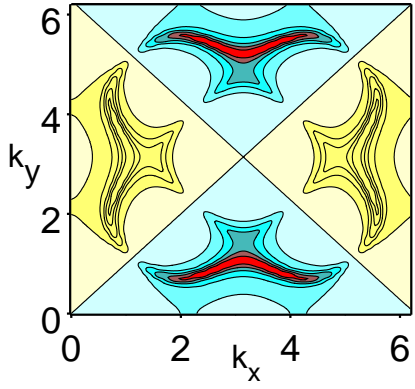}
\caption{(Color online) 2D projection of the superconducting gap
function $ \varphi({\bf k}) $ for $0 \leq k_x, k_y \leq 2\pi$.}
 \label{fig4}
\end{figure}

The   gap equation (\ref{10}) in strong-coupling approximation in
the imaginary Matsubara  frequency $\omega_{n}$  representation can be written
as,
\begin{eqnarray}
 \varphi({\bf k}) & = &
  \frac{T}{N}\sum_{\bf q} \, \sum_{n}\,
  \frac{{\varphi}({\bf q}) } {[Z_{\bf k}\,\omega_{n}]^2 +
 \varepsilon^2_{\bf q} }\;
     \nonumber \\
& \times & \left[ J_{\bf k-q} -  V_{\bf k-q} - |t({\bf q})|^{2}
\chi_{sf}({\bf k-q}) \right] \, ,
 \label{17}
 \end{eqnarray}
where $Z_{\bf k} = 1+ \lambda_{\bf k} =  1 -(d/ d\omega){\rm
Re}(\Sigma({\bf k},\omega)|_{\omega =0}$ is the quasiparticle
weight.  The latter is determined by the normal self-energy
(\ref{6}) which depends on contributions from all $l$-channels of
interactions expanded in a series of the Legendre polynomials
$P_l(\cos \Theta)$, contrary to the anomalous self-energy
(\ref{7}) where only the $l=2$ channel contributes to the
$d$-wave pairing. Therefore, a strong EPI in the $l=0$ channel
resulting in  a large effective mass renormalization, large $Z_{\bf k}$, is
unimportant for the $d$-wave pairing and can only suppress
the superconducting $T_c$ (see also~\cite{Lichtenstein95}).
Figure~\ref{fig3} shows doping dependence  $T_c(\delta)$ in units of
$t \sim 0.4$~eV for $Z_{\bf k} = 3$ where $T_c(\delta)$ induced by partial
contributions, AF and Coulomb interactions in MFA $\propto
(J_{\bf k-q} -  V_{\bf k-q})$ and spin fluctuations, $\propto
|t({\bf q})|^{2}$ are also shown. The maximal $T_c$ is of the order of $150$~K, while
for $Z_{\bf k} = 1$ its value  appears about five times higher. The
gap function found for the hole concentration  $\, \delta = 0.12 \,$ is shown in
Fig.~\ref{fig4} which clearly demonstrates the $d$-wave symmetry.
\par
In summary, we can conclude  that  the superconducting pairing
mediated by the  AF exchange interaction in MFA is  suppressed
by the screened Coulomb interaction and only charge fluctuations
cannot produce superconducting pairing as found in
Ref.~\cite{Alexandrov11}. However, spin-fluctuations induced by
the kinematic interaction give a substantial contribution to the
$d$-wave pairing and high-$T_c$ can be achieved.  EPI can be
important for the $d$-wave pairing only for  particular phonon
modes having a large $l=2$ component, while polaronic effects
induced by a large $l=0$ component of the EPI may be detrimental
for superconductivity in cuprates.

\begin{acknowledgements}
The results presented in the report have been obtained in
collaboration with A.\,A. Vladimirov and D. Ihle
(Sect.~\ref{sec:1}), and V.\,S. Oudovenko (Sect.~\ref{sec:2}).
\end{acknowledgements}

\end{document}